\documentclass[prl,twocolumn,showpacs,amsmath,amssymb,superscriptaddress]{revtex4-1}
\usepackage[english]{babel}
\usepackage{amsmath}
\usepackage{graphicx, nicefrac}
\usepackage{float}
\usepackage{siunitx}

\usepackage{dcolumn}
\usepackage{bm,graphicx}
\usepackage{etoolbox}

\usepackage[colorlinks]{hyperref}
\usepackage{url}
\usepackage[colorlinks]{hyperref}
\usepackage[normalem]{ulem}
\usepackage[dvipsnames,usenames]{color}
\usepackage{xcolor}
\usepackage[version=3]{mhchem} 

\begin{document}

\title{Addressing Shape and Extent of Weyl cones in TaAs by Landau level spectroscopy}

\author{D. Santos-Cottin}
\affiliation{Department of Physics, University of Fribourg, 1700 Fribourg, Switzerland}

\author{J.~Wyzula}
\affiliation{LNCMI, CNRS-UGA-UPS-INSA, 25, avenue des Martyrs, F-38042 Grenoble, France}

\author{F. Le Mardel\'e}
\affiliation{Department of Physics, University of Fribourg, 1700 Fribourg, Switzerland}

\author{I.~Crassee}
\affiliation{LNCMI, CNRS-UGA-UPS-INSA, 25, avenue des Martyrs, F-38042 Grenoble, France}

\author{E.~Martino}
\affiliation{Department of Physics, University of Fribourg, 1700 Fribourg, Switzerland}
\affiliation{IPHYS, EPFL, CH-1015 Lausanne, Switzerland}

\author{G.~Eguchi}
\affiliation{Institute of Solid State Physics, Vienna University of Technology, Wiedner Hauptstrasse 8-10, 1040 Vienna, Austria}

\author{Z.~Rukelj }
\affiliation{Department of Physics, University of Fribourg, 1700 Fribourg, Switzerland}
\affiliation{Department of Physics, Faculty of Science, University of Zagreb, Bijeni\v{c}ka 32, HR-10000 Zagreb, Croatia}
\author{M. Novak}
\affiliation{Department of Physics, Faculty of Science, University of Zagreb, Bijeni\v{c}ka 32, HR-10000 Zagreb, Croatia}

\author{M.~Orlita}
\affiliation{LNCMI, CNRS-UGA-UPS-INSA, 25, avenue des Martyrs, F-38042 Grenoble, France}
\affiliation{Institute of Physics, Charles University in Prague, CZ-12116 Prague, Czech Republic}
\author{Ana Akrap}
\email[]{ana.akrap@unifr.ch}
\affiliation{Department of Physics, University of Fribourg, 1700 Fribourg, Switzerland}

\date{\today}
\begin{abstract}
TaAs is a prime example of a topological semimetal with two types of Weyl nodes, W$_1$ and W$_2$, whose bulk signatures have proven elusive. We apply Landau level spectroscopy to crystals with multiple facets and identify---among other low-energy excitations between parabolic bands---the response of a cone extending  over a wide energy range. Comparison with density functional theory studies allows us to associate this conical band with nearly isotropic W$_2$ nodes. In contrast, W$_1$ cones, which are more anisotropic and less extended in energy, appear to be buried too deep beneath the Fermi level. They cannot be accessed directly. Instead, the excitations in their vicinity give rise to an optical response typical of a narrow-gap semiconductor rather than a Weyl semimetal.
\end{abstract}

\maketitle

%
%
Topological systems offer glimpses of relativistic-like physics within a crystal. 
The two most famous manifestations are Klein tunnelling in graphene \cite{Young2009}, and the chiral anomaly \cite{Armitage2018} in Weyl semimetals. These are materials whose broken inversion symmetry leads to pairs of Weyl nodes in the reciprocal space. The elementary excitations near those nodes are chiral Weyl quasiparticles.  
In TaAs---the first discovered and the most prominent Weyl semimetal---24 such nodes are sprinkled around the Brillouin zone \cite{Lv2015,Weng2015}. Among them, 8 are strongly anisotropic W$_1$ nodes, and 16 are fairly isotropic W$_2$ nodes \cite{Arnold2016}.
The large number and two different kinds of Weyl nodes have complicated our understanding of the underlying Weyl physics.
TaAs was proposed as a platform to benefit from the special properties of Weyl fermions \cite{Ma2017, Gao2020, Osterhoudt2019}.
Yet, most basic questions remain unanswered: How can one directly access the linearly dispersing Weyl cones? What is their energy range? When do their excitations dominate over other collective excitations?

We address the Weyl cones of TaAs through Landau level and infrared spectroscopy, measuring several samples with multiple crystalline facets. 
Our key finding is that only one kind of Weyl cones in TaAs significantly contributes to its optical response. These cones are isotropic and large, extending over more than 250 meV. Comparison with density functional theory (DFT) studies allows us to recognize them as W$_2$ cones.
In contrast, the W$_1$ cones are buried too deep below the Fermi energy to be optically accessible. Regardless, the electronic bands in their vicinity strongly contribute to the optical and magneto-optical response, because of their profound anisotropy, mirrored in a large joint density of states. This contribution dominates when the magnetic field is oriented along the [001] axis, effectively masking excitations in the W$_2$ cones.  

%
%
We grew single crystals of TaAs exposing large mirror-like surfaces in different crystalline planes: (001), (101), (112).
The samples were characterized through magneto-transport and quantum oscillations \cite{SM}.
Infrared spectra were then taken using FTIR Vertex 70v spectrometer and employing an \emph{in situ} coating technique \cite{Homes1993}.
Magneto-optical spectra were measured at 2 K, in fields up to 34~T.
A single sample with large (001), (101) and (112) planes allowed us to confirm that all the measurements on different samples are reproducible and consistent \cite{SM}.
%
%
%
%
\begin{figure}[!b]
	\includegraphics[width=0.9\linewidth]{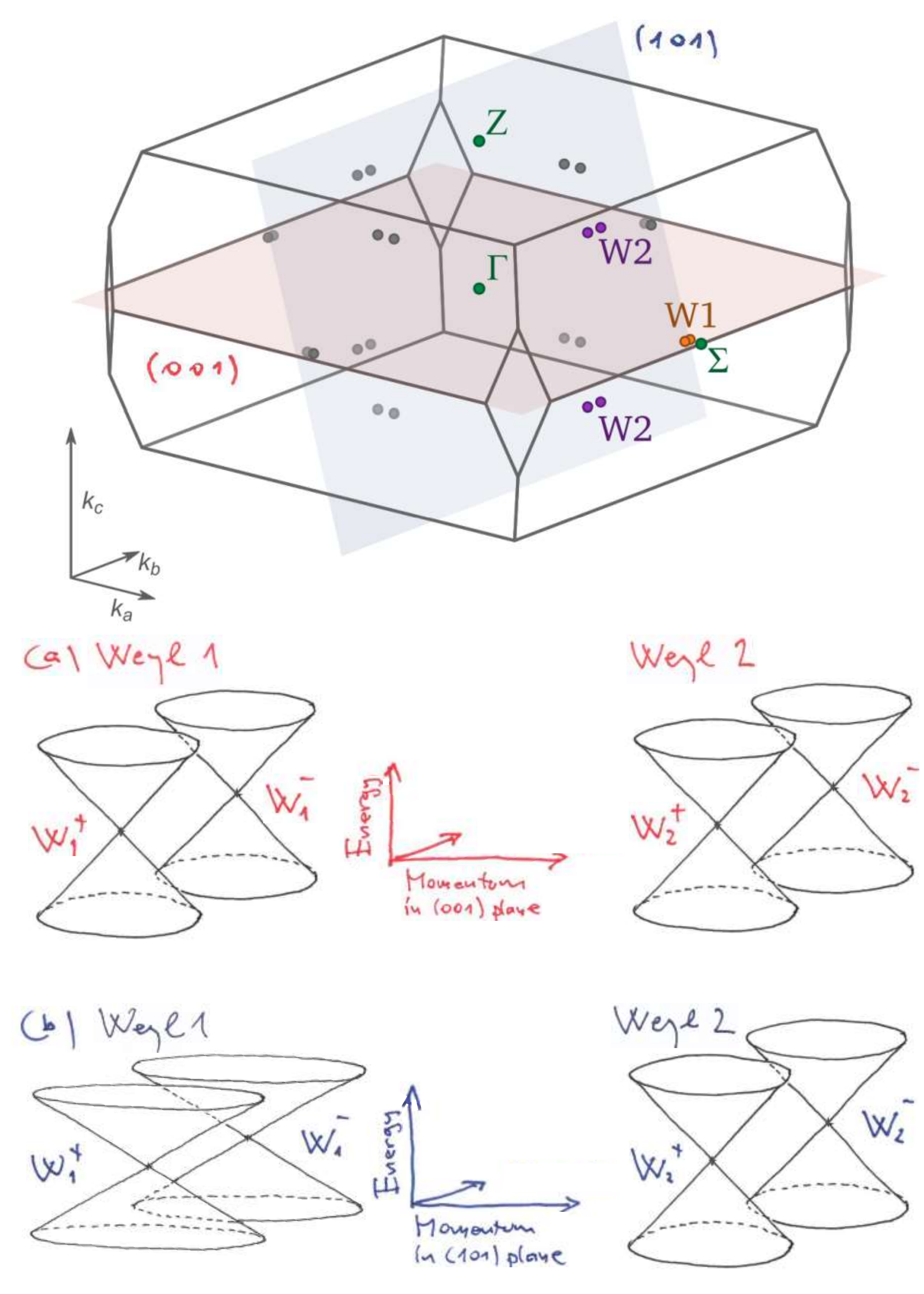}
	\caption{Top: Brillouin zone of TaAs, with (001) and (101) planes shown in red and blue, respectively. Following the band structure calculations \cite{Grassano2018}, we sketch the W$_1$ and W$_2$ Weyl cones
(a) in the (001) plane, and (b) in the (101) plane.	
	}
	\label{fig0}
\end{figure}
%
%

%
%
\begin{figure*}[!th]
	\includegraphics[width=0.95\linewidth]{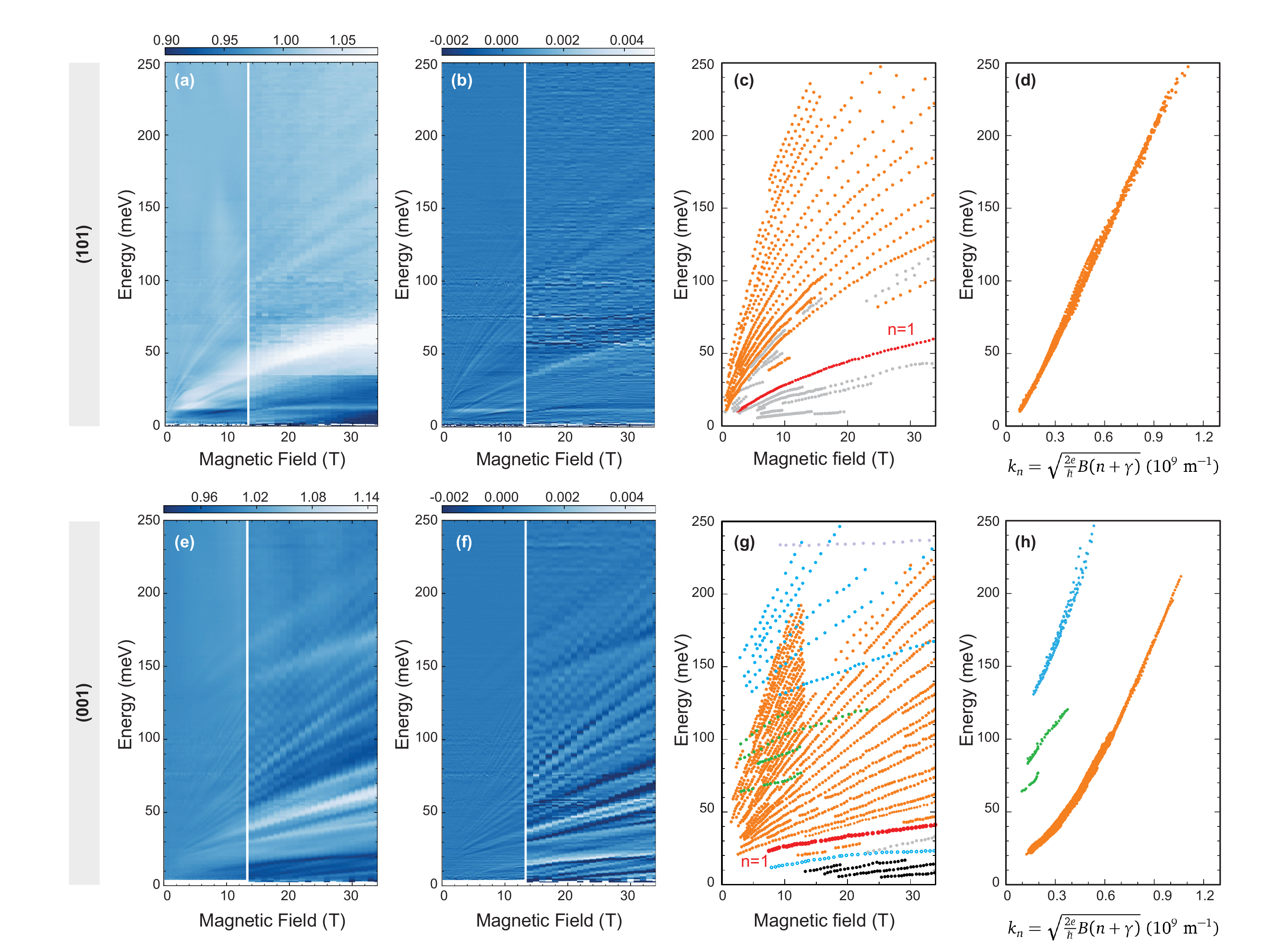}
	\caption{Inter Landau-level transitions in (101) (top) and (001) planes (bottom). Relative magneto-reflectance $R_B/R_0$ is shown in panels (a) and (e), and its derivative in (b) and (f). The inter-LL transitions are isolated in (c) and (g). They are plotted in (d) and (h) as a function of momentum $k_n=\sqrt{2eB(n+\gamma)/\hbar}$, with $\gamma=0$. 
	}
	\label{fig1}
\end{figure*}
%
%
%

%
%
The tetragonal symmetry of TaAs leads to a high number of Weyl points, shown in Fig.~\ref{fig0}.
Eight W$_1$ nodes with $k_c=0$ sit near the time-reversal invariant momentum point $\Sigma$ of the first Brillouin zone. Sixteen W$_2$ nodes lurk further away from high symmetry points, with $k_c \neq 0$. 
The connecting lines between all pairs of Weyl nodes are parallel with the (001) plane. 
The crystal unit cell is 4 times longer along the $c$ axis than along the $a$ axis. 
Such an elongated shape creates a steep angle of $\sim 74^\circ$ between the (001) and (101) planes, and an angle of $\sim 67^\circ$ between the planes (001) and (112).
The (001) plane of TaAs is most commonly addressed in the literature.
In contrast, other directions are much less explored. 
Figure~\ref{fig0}a schematically shows the shape of W$_1$ and W$_2$ cones in the (001) plane, after the DFT calculations \cite{Grassano2018}. A cut through the same Weyl cones is shown for the (101) plane in Fig.~\ref{fig0}b. Evidently, the W$_2$ cone is isotropic, but the W$_1$ cone becomes rather flat in the (101) plane.
This flat energy dispersion of the W$_1$ cone along the $c$ axis leads to an increased density of states (DOS). 
One may expect the same for the electronic states surrounding these cones.

Topological semimetals can be understood in fine detail using Landau level (LL) spectroscopy \cite{Orlita2014,Akrap2016,Hakl2017,Martino2019,Polatkan2020,Shao2019,Shao2020}. Their small Fermi pockets and high carrier mobility enable us to reach the quantum regime and observe the inter-LL transitions, letting us map the band structure. 
Figures \ref{fig1}a-b and  \ref{fig1}e-f show the colorplots of relative magneto-reflectance and its energy derivative, for (101) and (001) facet, respectively.
The observed inter-LL resonances are narrow, so we associate the positions of maxima in $R_B/R_0$ spectra directly with the transition energies.
The spectra were confirmed on different samples \cite{SM}, with similar carrier concentrations to the literature \cite{Arnold2016, Ramshaw2018}.
Magneto-optical spectra are exceptionally rich.
Far below 1~T, TaAs reaches its quantum regime and the inter-LL transitions become apparent. 
Figure~\ref{fig1}a-c shows a striking series of inter-LL transitions along the (101) facet. 
The spectra are dominated by a series of sublinear lines -- a nearly $\sqrt{B}$ dependence of interband excitations on magnetic field. This indicates a strongly non-parabolic dispersion,
as expected for a conical dispersion stemming from Weyl nodes \cite{AshbyPRB14,Polatkan2020}.

TaAs is a multiband system with a complex magneto-optical response, therefore we cannot analyze the experimental data using simple effective Hamiltonians \cite{Akrap2016,Hakl2017,Martino2019,Shao2019,Polatkan2020}.
Instead, to understand our spectra, we adopt a semiclassical approach \cite{Ashcroft}.
Lifshitz-Onsager formula says that in a magnetic field $B$, the area in momentum space encircled during cyclotron motion of an electron with crystal momentum $k_n$ equals $S_n = \pi k_n^2 = (eB/h) (n + \gamma)$. 
Here $n$ is an integer enumerating individual cyclotron orbits, in fact Landau levels. The parameter $\gamma$ describes the Berry phase of the explored band. 
In the semiclassical approach, we may assume that the optical excitations conserve the momentum of electrons which are promoted between the electronic bands. Each interband inter-LL transition, observed at a given $B$ and energy $\hbar\omega$, then measures the distance between bands at a given momentum $k_n=\sqrt{(n+\gamma)2eB/\hbar}$. In our case, rather than a Berry phase, $\gamma$  represents an additional tuning parameter, compensating for the fact that momentum is no longer a good quantum number in quantizing magnetic fields. In reality, the excited electrons change their LL index, or the cyclotron orbit. For a simple system with full rotational symmetry around the direction of $B$, the selection rules $n\rightarrow n\pm1$ apply in the Faraday configuration.     

The described procedure allows us to deduce the joint profile, $E_c-E_v$, of the involved conduction and valence bands. Assigning each excitation a particular orbital integer $n$, we obtain a fairly smooth profile (Fig. \ref{fig1}d), leading us to conclude that a well-defined conical band extends over more than 250~meV. 
There is no deviation from linearity up to highest experimentally probed energies. 
Assuming a full electron-hole symmetry, the slope gives a velocity of $v = (2.0\pm0.2)\times10^5$~m/s. 
This simple result of our analysis is surprising. We observe only one type of Weyl cones, while the other one does not manifest in our data. The occupation effect---Pauli blocking---is the most likely explanation. We may also conclude that the observed conical band is nearly isotropic. Otherwise, the pairs of Weyl cones oriented along $a$ and $b$ crystallographic axes would have to contribute differently to the magneto-optical response when $B$ is applied along the [101] direction, see Brillouin zone in Fig.~\ref{fig0}. A weak anisotropy of the traced conical bands may explain the appearance of additional, fairly weak lines in our data (gray points in Fig.~\ref{fig1}c). The analysis of the data collected on the (112)-oriented facet---$43^\circ$ away from the (101) plane---also indicates a single conical band with a nearly identical velocity parameter \cite{SM}. 
DFT studies \cite{Weng2015, Grassano2018} show that $W_2$ cones should be isotropic. In contrast, W$_1$ cones have a flat dispersion along the tetragonal $c$ axis, and their character is closer to  quasi-two-dimensional rather than three-dimensional. Therefore, the conical feature we observe is W$_2$. 
 
In stark contrast, the (001) data in Fig.~\ref{fig1}e--h shows over 50 discernible lines, where linear-in-$B$ transitions completely dominate the response. Clearly, in the (001) plane we are primarily looking at excitations between parabolic-like bands. 
No trace is seen of $\sqrt{B}$-like transitions, which would be primarily expected in a 3D Weyl semimetal.
We may identify at least four separate series of transitions, plotted in different colours in Fig.~\ref{fig1}g. 
The oscillator strengths of these transitions are considerably stronger than for the excitations observed on other facets \cite{SM}, effectively masking the $\sqrt{B}$-like transitions. 
At the same time, these linear-in-$B$ excitations are completely missing from the response of other facets. This apparent anisotropy allows us to establish that these excitations originate in the vicinity of the W$_1$ cones---around the $\Sigma$ point---where the band structure becomes significantly flatter in the tetragonal $c$ direction \cite{Grassano2018}. 

Let us now analyze the response of the (001) facet in greater detail. 
Repeating the semiclassical approach \cite{Miyata2017}, we first identify which transitions belong together. 
The energy of each inter-LL transition is then plotted as a function of $k_n$, after assigning each relevant transition an index $n$.
We recognize up to four families of inter-LL transitions on the (001) facet. 
The most pronounced series of lines extrapolate to 15~meV in the limit of vanishing magnetic fields.  
More than 15 transitions collapse onto the same line in the energy versus $k_n$ plot in Fig. \ref{fig1}h.
From this ``collapsed'' plot, representing joint bands, we see that the underlying nearly parabolic bands have a gap of $\Delta \sim 15$~meV.
A vertical line at any high field in Fig.~\ref{fig1}g  intersects a series of equidistant excitations. Their spacing, amounting to 15--20 meV at 30~T, gives the effective mass of these carriers, $m^* \approx 0.4 m_e$.
The lowest transition, 0$\rightarrow$1, emerges at 6--8 T, and is marked in dark red in Fig.~\ref{fig1}g.
This value gives us the size of the relevant Fermi pocket. In quantum oscillations, we have a frequency of 7--7.5 T, a size of the W$_1$ pocket which corresponds well to previously reported values \cite{Arnold2016,Ramshaw2018, SM}.
Such a small pocket size would imply that W$_1$ is in the chiral regime. Yet, 
the response in the (001) plane appears to be dictated by a narrow-gap parabolic band. 
This striking absence of any observable Weyl cone signatures in the (001) orientation is an important result. 
It tells us that the W$_1$ cones are inaccessible for optical excitations, despite a very low carrier concentration. 

TaAs is closely related to TaP, whose (001) plane shows peculiar down-dispersing inter-LL transitions \cite{Polatkan2020}. Those transitions are related to the band inversion and the linear dispersion from W$_1$ cones.
In TaAs, nothing similar can be seen: no lines shift to lower energies as the field increases.
This is another strong indication that the Fermi level in TaAs is not placed within the W$_1$ cones, and the system is above the topological separatrix. 
Finally, on the (001) facet of TaAs we see no evidence of non-parabolic behavior at high energies, contrary to  TaP \cite{Polatkan2020}. 
In TaAs, the W$_1$ nodes do not seem to play a role, and the conical dispersion from W$_2$ points becomes remarkably evident once we move away from the (001) plane.

Fig.~\ref{fig1}g contains several unusual low-energy lines shown in black.
These lines are only seen in high fields.
The lowest line begins at 25~T, crosses 7 meV at 30 T, and extrapolates to zero energy at zero field.
Assuming this is a cyclotron resonance (CR), its associated effective mass is $m^*_{H} = 0.5 m_e$. 
Such heavy carriers in TaAs may be the trivial holes seen in quantum oscillations \cite{Arnold2016,SM}. 
The CR line onset at 25~T corresponds to the heavy-hole pocket quantum limit. 
Three roughly equidistant lines appear above the lowest line.
One may speculate that these are CR harmonics appearing due to strongly anisotropic band dispersion, similar to bulk graphite with electrons in the vicinity of a Lifshitz transition \cite{Orlita2012}.
The trivial hole pocket in TaAs is indeed anisotropic and warped \cite{Arnold2016}.

One pronounced yet unassigned inter-LL transition in the (001) plane is the line indicated in blue in Fig.~\ref{fig1}g. 
This line does not naturally belong with the straight lines emanating from 15 meV. Instead, it extrapolates linearly into zero energy at zero field, but flattens significantly above 20~T. Such response may represent CR absorption of massive electrons in one of partly occupied, strongly non-parabolic bands.   

%
%
\begin{figure}[!h]
	\includegraphics[width=0.9\linewidth]{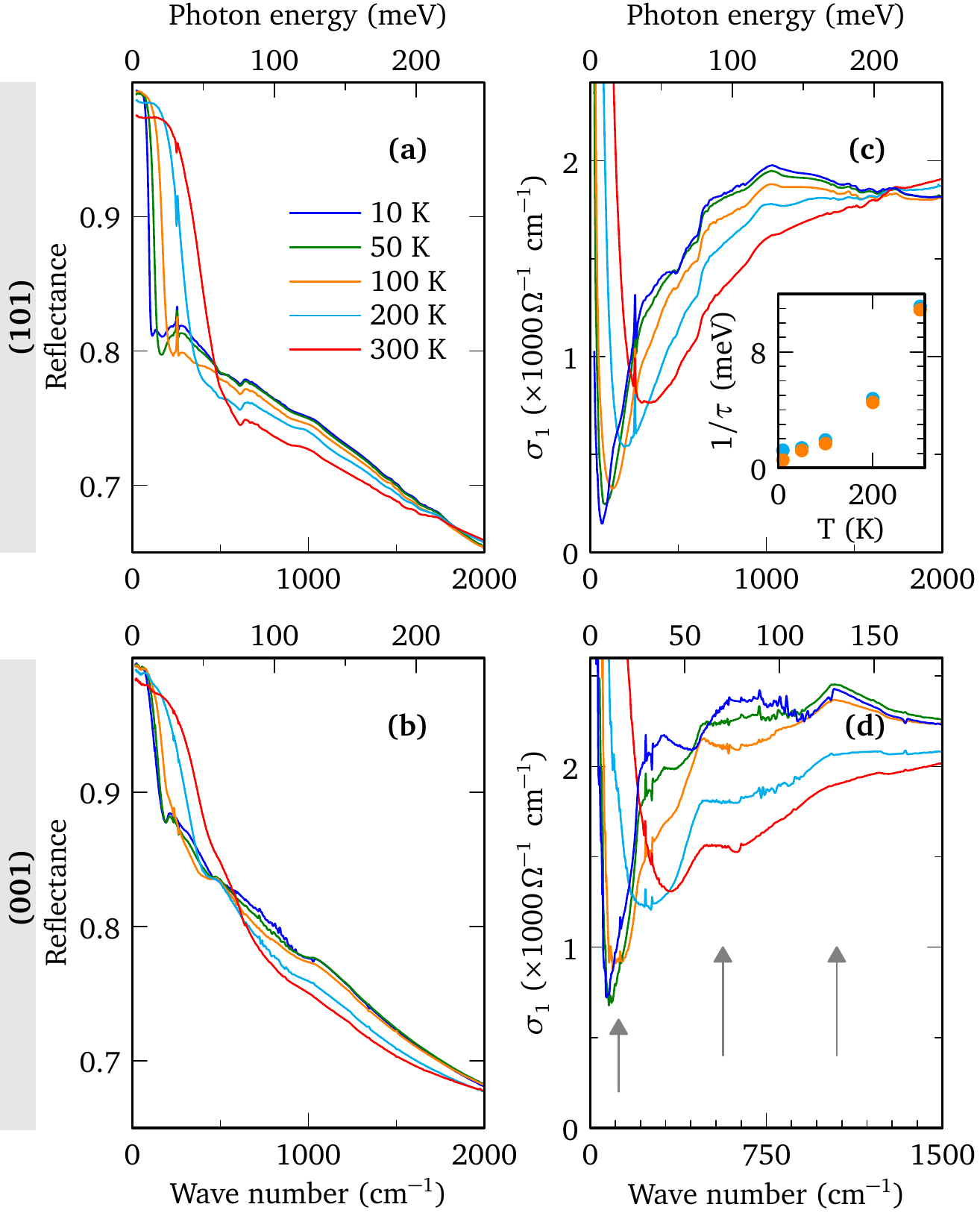}
	\caption{Zero-field optical properties of TaAs, for (101) (top) and (001) plane (bottom panels).
	Far-infrared reflectance is shown in (a) and (b) as a function of photon energy, at different temperatures.
	The real part of optical conductivity, $\sigma_1$, is shown in (c) and (d). 
	Grey arrows indicate the 15 meV absorption onset, and van Hove peaks at 70~meV and 128~meV.
	Inset shows the temperature dependence of scattering rate $1/\tau$ for the two planes.
	}
	\label{fig2}
\end{figure}

For both (101) and (001) facets, LL spectroscopy paints a full and complex picture of low-energy excitations in TaAs.
This knowledge helps us better understand its zero-field infrared spectra. Optical conductivity primarily reflects the joint density of states, and thereby becomes strongly sensitive to the out-of-plane dispersion.
Figure~\ref{fig2} shows the zero-field optical spectra of TaAs for (101) and (001) planes.
The low-temperature reflectivity shows sharp plasma edges at very low energies (15--20~meV), in line with low carrier concentration and small scattering rate $1/\tau$, which are necessary conditions to observe the LL quantization in fields below 1~T (Fig.~\ref{fig1}a,e). 
Through Kramers-Kronig transformations, we obtain the complex optical conductivity, $\hat{\sigma}=\sigma_1+ i\sigma_2$.
Its real part $\sigma_1(\omega)$ is shown in Fig.~\ref{fig2}c-d. 
The large carrier mobility is reflected in a narrow Drude response, with $1/\tau \sim 1$~meV. Insets in Fig.~\ref{fig2} shows that $1/\tau$ follows a quadratic temperature dependence.

Most relevant to the Weyl node physics are the interband excitations, which in TaAs set in at remarkably low energies. 
The 10~K absorption onset is at 15~meV for the (001) facet, 
in agreement with the gap from Fig.~\ref{fig1}h. 
Interestingly, $\sigma_1(\omega)$ has sharper features for the (001) facet, Fig.~\ref{fig2}d, than for the (101) facet, Fig.~\ref{fig2}c.
This is caused by the flat dispersion of the bands near $\Sigma$ point \cite{Grassano2018}, enhancing their optical response.
Contrary to initial reports \cite{Xu2016a, Kimura2017}, we see several sharp kinks in $\sigma_1(\omega)$. 
Besides the absorption onset kink at 15~meV, the peaks at 70 and 128~meV can be linked to the zero-field extrapolation of several lines in Fig.~\ref{fig1}h.
These may be van Hove singularities that could originate anywhere in the Brillouin zone, but because of the DOS effect, their source is most likely in the bands near the $\Sigma$ point.
All those fine features are superimposed upon a broad background. 
We find that there is no simple meaning of the slopes in $\sigma_1(\omega)$, nor a link to Weyl cone effective velocities. The W$_1$ pocket, which dominates the (001) plane response, is not in its chiral limit.
In the (101) plane, Fig~\ref{fig2}c, $\sigma_1(\omega)$ has a smoother profile, with a narrow Drude component and a broad maximum at 120~meV. We expect a contribution of W$_2$ nodes in this orientation, although it is difficult to say what this contribution should look like. 
It seems abundantly clear that the W$_2$ contribution is not simply a straight line from 0 to the maximum cone extent.

In conclusion, we found a strong difference in the optical and magneto-optical response of TaAs for different orientations, specifically planes (001) and (101). This anisotropic response is caused by the anisotropy of bands giving rise to the W$_1$ nodes.
Crucially, we found that W$_1$ nodes are not optically accessible even in low-carrier-density crystals of TaAs. The magneto-optical response of the (001) facet, most often addressed experimentally, does not show any direct signatures of Weyl physics. 
Instead, optical experiments can only probe the neighboring parabolic bands. 
In contrast, our magneto-optical data give strong evidence that W$_2$ cones are highly isotropic and extend over more than 250~meV. They are characterized by a relatively low velocity parameter of $(2.0\pm0.2)\times10^5$~m/s.
Therefore, in the (101) plane and other planes closer to the crystal $c$ axis, optical experiments can readily detect topologically nontrivial bands.

%
%
We thank W. Bechstedt for deeply informative discussions, and N. Miller for helpful comments.
A.A. acknowledges funding from the  Swiss National Science Foundation through project PP00P2\_170544.
M.N. acknowledges support of CeNIKS project co-financed by the Croatian Government and the EU through the European Regional Development Fund Competitiveness and Cohesion Operational Program (Grant No. KK.01.1.1.02.0013).
This work has been supported by the ANR DIRAC3D. We acknowledge the support of LNCMI-CNRS, a member of the European Magnetic Field Laboratory (EMFL).

\bibliography{TaAs}

\end{document}